\documentclass[a4paper,aps,prd,showpacs,showkeys,amsmath,amssymb,amsfonts,nofootinbib,floats,floatfix,superscriptaddress,eqsecnum]{revtex4}

\usepackage[pdftex,breaklinks=false,bookmarksnumbered]{hyperref}

\def\dd{{\mathrm{d}}}

\def\vct#1{\mathbf{#1}}
\def\myrowsep{\\ [0.7ex]}

\def\pacsn{04.25.Nx, 04.20.Fy}
\def\keyw{Post-Newtonian approximation, classical spin, canonical formalism, Effective Field Theory.}

\begin{document}

%
%

\title{On the comparison of results regarding the post-Newtonian approximate treatment of the dynamics of extended spinning compact binaries}
 \author{Steven Hergt}
\email{steven.hergt@uni-jena.de}
\affiliation{Theoretisch--Physikalisches Institut, \\
	Friedrich--Schiller--Universit\"at Jena, \\
	Max--Wien--Platz 1, 07743 Jena, Germany, EU}

 \author{Jan Steinhoff}
\email{jan.steinhoff@uni-jena.de}
 \affiliation{Theoretisch--Physikalisches Institut, \\
	Friedrich--Schiller--Universit\"at Jena, \\
	Max--Wien--Platz 1, 07743 Jena, Germany, EU}
 \affiliation{Centro Multidisciplinar de Astrof\'isica (CENTRA), Departamento de F\'isica, \\
	Instituto Superior T\'ecnico (IST), Universidade T\'ecnica de Lisboa, \\ 
	Avenida Rovisco Pais 1, 1049-001 Lisboa, Portugal, EU}

\author{Gerhard Sch\"afer}
\email{gos@tpi.uni-jena.de}
\affiliation{Theoretisch--Physikalisches Institut, \\
	Friedrich--Schiller--Universit\"at Jena, \\
	Max--Wien--Platz 1, 07743 Jena, Germany, EU}
\pacs{\pacsn}
\keywords{\keyw}

\begin{abstract}
A brief review is given of all the Hamiltonians and effective potentials calculated hitherto covering the post-Newtonian (pN) dynamics of a two body system.
A method is presented to compare (conservative) reduced Hamiltonians with nonreduced potentials directly at least up to the next-to-leading-pN order.
\end{abstract}

\date{\today}

\maketitle

\section{Post-Newtonian modeling and results of the two body dynamics}
The post-Newtonian (pN) treatment of the dynamics of two bodies in general relativity has to incorporate both spin and tidal force induced mass multipoles
of the constituents of the physical systems \cite{Binnington:Poisson:2009}. In our analysis we will focus on the spin multipole degrees of freedom and present some
illuminating results, that will be useful for future research of gravitational wave date extraction from such a system. One easy way to start
to model such a system is by choosing the representation by Tulczyjew's singular stress-energy tensor\cite{Tulczyjew:1959} $T_{\mu\nu}$, with the greek coordinate indices running from 0 to 3,
in the following way
\begin{eqnarray}
\sqrt{-g} T^{\mu\nu}(x^{\sigma})& =& \int d \tau \bigg[
	u^{(\mu} p^{\nu)} \delta_{(4)}+\left(u^{(\mu}S^{\nu)\alpha}\delta_{(4)}\right)_{||\alpha}+\frac{1}{3}R_{\alpha\beta\rho}^{(\mu}J^{\nu)\rho\beta\alpha}\delta_{(4)}\nonumber\\
                                &&\quad-\frac{2}{3}\left(J^{\mu\alpha\beta\nu}\delta_{(4)}\right)_{||(\alpha\beta)}+\dots\bigg]
\,,\quad u^{\mu} = \frac{d z^{\mu}}{d \tau}\,, \quad \delta_{(4)} = \delta(z^{\sigma} - x^{\sigma})
\end{eqnarray}
with the body's 4-velocity $u^{\mu}$, the 4-momentum $p_{\mu}$, the antisymmetric spin tensor $S^{\mu\nu}$ modeling the pole-dipole structure and Dixon's reduced quadrupole moment tensor
$J^{\mu\alpha\beta\nu}$ modeling first order finite size effects while possessing the same symmetries as the Riemann tensor $R^{\mu\alpha\beta\nu}$. It follows a decomposition of
$J^{\mu\alpha\beta\nu}$ into stress, flow and the symmetric trace-free mass quadrupole $Q^{}_{\mu\nu}$. The latter is given by the ansatz with a vector $f_{\mu}$ to which the spin
is orthogonal, $S^{\mu\nu}f_{\nu}=0$
\begin{equation}
 Q_{\mu\nu}=\frac{C_{Q}}{m}\left(S_{\mu\rho}S_{\nu}^{\;\;\rho}-\frac{1}{3}P_{\mu\nu}S^{\rho\sigma}S_{\rho\sigma}\right)\,,\quad P^{\mu\nu}=g^{\mu\nu}-\frac{1}{f_{\rho}f^{\rho}}f^{\mu}f^{\nu}\,.
\end{equation}
and is parametrized only by $C_{Q}$ in the Newtonian limit and quadratic level in spin fully encoding the rotational deformation. For black holes one has
 $C_{Q}=1$ \cite{Thorne:1980} while for neutron star models $C_{Q}$ depends on the equations of state \cite{Laarakkers:Poisson:1999} and varies between $4.3\dots7.4$.
The next step to perform explicit pN calculation is complex in various ways. We compare two prominent methods. One method aims at
calculating a Hamiltonian. This is achieved by a 3+1 decomposition of Einstein's field equations and the energy-momentum tensor from Eq. (1) leading to constraints which have to be fulfilled
at all times on the 3-dimensional hypersurfaces orthogonal to the time direction. We then 
use the ADM formalism as outlined in \cite{Steinhoff:2011} to find the canonical set of variables ($\hat{\vct{z}}_{I}, \hat{\vct{p}}_{I},\hat{\vct{S}}_{I}$) with the body label $I=1,2$ fulfilling
their standard canonical Poisson bracket relations $\{\hat{z}^i_{I},\hat{p}_{Jj}\}=\delta_{ij}\delta_{IJ}$ and $\{\hat{S}_{I(i)},\hat{S}_{I(j)}\}=\epsilon_{ijk}\hat{S}_{I(k)}$
 with $i,j,k$ running from 1 to 3 and the round brackets around them indicate the components of local Lorentz indices $a,b,\dots$
 from the beginning of the alphabet, so $a\in\{(0),(i)\}$. The spin tensor $S_{ab}$ defined in a local Lorentz frame is therefore connected to the coordinate frame by a vierbein
 transformation $S_{ab}=e_{a\mu}e_{b\nu}S^{\mu\nu}$. The ADM formalism also leads to a formula for calculating the Hamiltonian in full reduced phase space 
 by imposing the ADM$TT$ or transverse traceless gauge to the 3-metric on the 3-hypersurface
 and by choosing the correct (canonical) spin supplementary condition (SSC) which fixes the center of the object.
 By expansion of the constraints in pN powers of $\frac{v^2}{c^2}\sim \frac{Gm}{rc^2}$ one ends up
 with a perturbative scheme to calculate Hamiltonians to formally arbitrary pN orders. The general Hamiltonian being the generator for the equations of motion of the binary therefore
intrinsically adopts the post-Newtonians expansion of the field equations. As the spin is of pN order $1/c$ or $1/c^2$ depending on its strenght the formal labeling is such that we call the first
post-Newtonian spin Hamiltonians not 1.5pN or 2.5pN according to formal counting rules but just the leading order (LO)
 ones and the higher corrections we call next-to-leading (NLO) and next-to-next-to-leading order (N$^2$LO). In table \ref{tab1} we give a list of all known pN Hamiltonians for the case of
 maximally rotating objects where $|\vct{S}|=\frac{Gm^2\chi}{c}$ with the dimensionless spin $\chi=1$.
\begin{table}
 \caption{\label{tab1}Post-Newtonian Hamiltonians known to date}
\begin{center}
\begin{tabular}{r@{}r*{15}{@{$\;$}c}}
order	&& 1.0 && 1.5 && 2.0 && 2.5 && 3.0 && 3.5 && 4.0 && 4.5 \myrowsep
& $H^{\text{N}}$ \myrowsep
PM	& $+$ & $H^{\text{1PN}}$ &&& $+$ & $H^{\text{2PN}}$ & $+$ & $H^{\text{2.5PN}}$ & $+$ & $H^{\text{3PN}}$ & $+$ & $H^{\text{3.5PN}}$ & $+$ & $(H^{\text{4PN}})$ & $+$ & $\{H^{\text{4.5PN}}\}$ \myrowsep
SO	&&& $+$ & $H^{\text{LO}}_{\text{SO}}$ && & $+$ & $H^{\text{NLO}}_{\text{SO}}$ && & $+$ & $H^{\text{N$^2$LO}}_{\text{SO}}$ & $+$ & $H^{\text{LO,R}}_{\text{SO}}$ & $+$ & $(H^{\text{N$^3$LO}}_{\text{SO}})$ \myrowsep
S$_1^2$		&&&&& $+$ & $H^{\text{LO}}_{\text{S$_1^2$}}$ && & $+$ & $H^{\text{NLO}}_{\text{S$_1^2$}}$ && & $+$ & $(H^{\text{N$^2$LO}}_{\text{S$_1^2$}})$ & $+$ & $\{H^{\text{LO,R}}_{\text{S$_1^2$}}\}$ \myrowsep
S$_1$S$_2$	&&&&& $+$ & $H^{\text{LO}}_{\text{S$_1$S$_2$}}$ && & $+$ & $H^{\text{NLO}}_{\text{S$_1$S$_2$}}$ && & $+$ & $H^{\text{N$^2$LO}}_{\text{S$_1$S$_2$}}$ & $+$ & $H^{\text{LO,R}}_{\text{S$_1$S$_2$}}$ \myrowsep
spin$^3$	&&&&&&&&&&& $+$ & $[H_{S^3}^{\text{LO}}]$ & && $+$ & $(H_{S^3}^{\text{NLO}})$ \myrowsep
spin$^4$	&&&&&&&&&&&&& $+$ & $[H_{S^4}^{\text{LO}}]$ && \\
$\vdots$		&&&&&&&&&&&&&&&& $\ddots$ \\
\end{tabular}
\begin{center}
 $\{.\}$ EOM known\hspace{0.5cm}$[.]$ for Black Holes only\hspace{0.5cm} $(.)$ not known (yet)
\end{center}
\end{center}
\end{table}
$H^{N}$ is the Newtonian Hamiltonian, PM means point mass, $H^{nPN}$ with $n\in\{1,2,\dots\}$ are the conservative pure point mass Hamiltonians, $H^{\frac{n}{2}PN}$
are the dissipative (radiative) pure point mass Hamiltonians, SO refers to spin-orbit
coupling, $S_{1}S_{2}$ to spin(1)-spin(2) coupling and $S_{1}^2$ to spin quadrupole coupling involving the constant $C_{Q}$.
 LO,R in the index indicates the dissipative counterpart to leading order, so $H_{S_{1}^2}^{LO,R}$ is the
radiative counterpart to the conservative part $H_{S_{1}^2}^{LO}$. Obviously, the radiative part is much higher
 in pN order then the conservative part, but nevertheless they are important
to cover the dynamics to 4.5pN order consistently; up until now the radiation field is known to 2.5pN order only.
One other method to arrive at pN equations of motion is the derivation of effective potentials which are subtly
 related to Hamiltonians by a Legendre transformation. This
derivation is most effectivly achieved by sophisticated methods from Effective Field Theory (EFT) that uses full knowledge
 from quantum field theoretical calculations.
 Up until now, pN potentials have been calculated to 3pN order \cite{Foffa:Sturani:2011} for point masses and to NNLO for spin(1)-spin(2) coupling \cite{Levi:2011}.

\section{Comparison between Effective Field Theory potentials and ADM Hamiltonians}

Effective potentials are part of a Lagrangian with the Newtonian kinetic energy $T_{N}$
\begin{equation}
 L_{eff}=T_{N}-V_{eff}=\frac{m_{1}}{2}v_{1}^2+\frac{m_{2}}{2}v_{2}^2-V_{eff}.
\end{equation}
The conservative effective potential $V_{eff}$ for two interacting bodies is pN expanded up to next-to-leading order (NLO) spin effects in the following way
\begin{equation}
 V_{eff} = V_{PM}
	+ V^{LO}_{SO} + V^{LO}_{S_1^2}
		+ V^{LO}_{S_2^2} + V^{LO}_{S_1S_2}
	+ V^{NLO}_{SO} + V^{NLO}_{S_1^2}
		+ V^{NLO}_{S_2^2}
		+ V^{NLO}_{S_1S_2}\,.
\end{equation}
One key difference between EFT potentials and ADM Hamiltonians is that in most cases the potentials still depend on the $S^{(0)(i)}$-components of the spin
tensor, which have to be fixed by 
choosing an appropriate SSC.
For a direct comparison a formal Legendre transformation of the nonreduced potentials is conducted yielding the effective Hamiltonian $H_{eff}$,
 which is been followed by a reduction process in phase space in order to arrive at a canonical set
 of variables, see \cite{Hergt:Steinhoff:Schafer:2011} for details.
This `canonicalization' is most transparently accomplished by reducing the following effective action
\begin{equation}
 S_{eff}=\int\dd t\, L_{eff}=\int\dd t\left(p_{1i}\dot z^{i}_{1}+p_{2i}\dot z^{i}_{2}-\frac{1}{2}S_{1ab}\Omega_{1}^{ab}-\frac{1}{2}S_{2ab}\Omega_{2}^{ab}
-H_{eff}\left(\vct{z}_{I},\vct{p}_{I},S_{Iab}\right)\right)\,.
\end{equation}
Here we have defined the angular velocity tensor $\Omega^{ab}\equiv\Lambda_{A}^{\;\;a}\dot{\Lambda}^{Ab}$ rendering $\Omega^{ab}$ antisymmetric and $\Lambda_{A\mu}\Lambda^{A}_{\;\;\nu}=g_{\mu\nu}$,
 $\Lambda_{Aa}\Lambda^{A}_{\;\;b}=\eta_{ab}$ with $(A,B,\dots)\in\{[0],[i]\}$ being the body-fixed frame labels. The reduced action has to read
\begin{equation}
 \hat{S}_{eff}=\int\dd t\left(\hat{p}_{1i}\dot{\hat{z}}^{i}_{1}+\hat{p}_{2i}\dot {\hat{z}}^{i}_{2}-\frac{1}{2}\hat{S}_{1(i)(j)}
\hat{\Omega}_{1}^{(i)(j)}-\frac{1}{2}\hat{S}_{2(i)(j)}\hat{\Omega}_{2}^{(i)(j)}-H_{can}\left(\hat{\vct{z}}_{I},\hat{\vct{p}}_{I},\hat{\vct{S}}_{I}\right)\right)
\end{equation}
with $\hat{\Omega}^{(i)(j)}=\hat{\Lambda}_{[k]}^{\;\;\;(i)}\dot{\hat{\Lambda}}^{[k](j)}$ given by a nonlinear shift of $\Lambda^{[k](i)}$ to $\hat{\Lambda}^{[k](i)}$
 so that $\hat{\Lambda}^{[k](i)}\hat{\Lambda}^{[k](j)}=\delta_{ij}$.
This reduction is achieved by inserting the covariant SSC $S_{ab}u^b=0$ as well as its conjugate condition $\Lambda^{[i]a}u_{a}=0$ into $\frac{1}{2}S_{ab}\Omega^{ab}$ and performing a pN
approximate variable transformation of spin and position reading
\begin{eqnarray}
 z_{1}^{i}&=&~\hat{z}_{1}^{i}-\bigg[\frac{1}{2m_{1}^2}p_{1k}\hat{S}_{1(i)(k)}\left(1-\frac{\vct{p}_{1}^2}{4m_{1}^2}\right)
-G\frac{m_{2}}{m_{1}^2}\frac{p_{1k}\hat{S}_{1(i)(k)}}{\hat{r}_{12}}+\frac{3}{2}G\frac{p_{2k}\hat{S}_{1(i)(k)}}{m_{1}\hat{r}_{12}}\nonumber\\
&&+\frac{G}{2}\frac{\hat{n}_{12}^{k}(\vct{\hat{n}}_{12}\cdot\vct{p}_{2})\hat{S}_{1(i)(k)}}{m_{1}\hat{r}_{12}}+G\frac{m_{2}}{m_{1}^2}\frac{\hat{S}_{1(k)(l)}\hat{S}_{1(i)(l)}\hat{n}_{12}^{k}}{\hat{r}_{12}^2}+G\frac{\hat{n}_{12}^{k}
\hat{S}_{1(i)(l)}\hat{S}_{2(k)(l)}}{m_{1}\hat{r}_{12}^2}\bigg]\,,
\end{eqnarray}
\begin{eqnarray}
 S_{1(i)(j)}&=&~\hat{S}_{1(i)(j)}-\bigg[\frac{p_{1[i}\hat{S}_{1(j)](k)}p_{1k}}{m_{1}^2}\left(1-\frac{\vct{p}_{1}^2}{4m_{1}^2}\right)-\frac{2Gm_{2}}{m_{1}^2\hat{r}_{12}}p_{1[i}\hat{S}_{1(j)](k)}p_{1k}\nonumber\\
 &&+\frac{3G}{m_{1}\hat{r}_{12}}p_{1[i}\hat{S}_{1(j)](k)}p_{2k}+\frac{G}{m_{1}\hat{r}_{12}}p_{1[i}\hat{S}_{1(j)](k)}\hat{n}_{12}^{k}(\vct{\hat{n}}_{12}\cdot\vct{p}_{2})\\
 &&+\frac{2Gm_{2}}{m_{1}^2\hat{r}_{12}^2}p_{1[i}\hat{S}_{1(j)](l)}\hat{S}_{1(k)(l)}\hat{n}_{12}^{k}+\frac{2G}{m_{1}
\hat{r}_{12}^2}p_{1[i}\hat{S}_{1(j)](l)}\hat{S}_{2(k)(l)}\hat{n}_{12}^{k}\bigg]\,.\nonumber
\end{eqnarray}
 Those formulas are valid to transform the potentials at least to NLO to their canonical Hamiltonian counterpart, which enabled us to obtain an overall agreement of all EFT NLO potentials
with their corresponding ADM Hamiltonian as displayed in table \ref{tab2} up to canonical transformations indicated by $\approx$, see \cite{Hergt:Steinhoff:Schafer:2011} for a thorough investigation.
\begin{table}
\caption{\label{tab2} Agreement between EFT potentials and ADM Hamiltonians}
\begin{minipage}[bl!]{0.5\textwidth}
\begin{tabular}{lllcl} 
 & & & & \\
$ V_{NLO}^{SO}$ & Levi\cite{Levi:2010} & & &\\& & & $H_{NLO ADM}^{SO}$ & Damour/Jaranowski/Sch\"afer\cite{Damour:Jaranowski:Schafer:2008:1,Steinhoff:Schafer:Hergt:2008}\\
$ V_{NLO}^{SO}$ & Porto\cite{Porto:2010} & & &\\& & & &\\\cline{1-2}\cline{4-5}& & $\approx$ & &\\
$ V_{NLO}^{S_{1}S_{2}}$ & Porto/Rothstein\cite{Porto:Rothstein:2008:1,Porto:Rothstein:2008:1:err} & & $ H_{NLO ADM}^{S_{1}S_{2}}$ & Steinhoff/Hergt/Sch\"afer\cite{Steinhoff:Hergt:Schafer:2008:2,Steinhoff:Schafer:Hergt:2008}\\& & & &\\\cline{1-2}\cline{4-5}& & & &\\
$ V_{NLO}^{S_{1}^2}$ & Porto/Rothstein\cite{Porto:Rothstein:2008:2,Porto:Rothstein:2008:2:err} & & $ H_{NLO ADM}^{S_{1}^2}$ & Hergt/Steinhoff/Sch\"afer
\cite{Hergt:Steinhoff:Schafer:2010:1}\\& & & &\\
\end{tabular}
\end{minipage}
\end{table}

\section*{Acknowledgments}
This work is supported by the Deutsche Forschungsgemeinschaft (DFG) through
SFB/TR7 ``Gravitational Wave Astronomy,'' project STE 2017/1-1, and GRK 1523,
and by the FCT (Portugal) through PTDC project CTEAST/098034/2008.

\section*{References}
\providecommand{\href}[2]{#2}\begingroup\raggedright

\endgroup

\end{document}